\documentclass[11pt]{article}
\renewcommand{\footnoterule}{%
  \kern -3pt
  \hrule width \textwidth height 0.5pt
  \kern 2pt
}

\usepackage{latexsym}
\usepackage{amssymb}
\usepackage{amsmath}
\usepackage{graphicx}
\usepackage{bm}
\usepackage{enumerate}
\usepackage{sectsty}
\usepackage{subfig}
\usepackage{float}
\usepackage{caption}
\usepackage{times,fancyhdr}
\usepackage{dsfont}

\sectionfont{\normalsize}
\subsectionfont{\small}

\numberwithin{equation}{section}

\newcommand*{\Scale}[2][4]{\scalebox{#1}{$#2$}}%
\newcommand{\ps}{phase space }
\newcommand{\iothree}{\mathcal{I}O(3)}
\newcommand{\iothreeplus}{\mathcal{I}O^{{\vspace{-0.9mm}\mbox{\tiny{$+$}}}}\hspace{-0.6mm}(3)}
\newcommand{\circled}{\overset{\circ}{\Delta}}
\newcommand{\sharpd}{\Delta^{\#}}
\newcommand{\wn}{W(n^{1},\,n^{2},\,n^{3})}
\newcommand{\vx}{V(x^{1},\,x^{2},\,x^{3})}
\newcommand{\wzn}{W_{(0)}(n^{1},\,n^{2},\,n^{3})}
\newcommand{\vzx}{V_{(0)}(x^{1},\,x^{2},\,x^{3})}
\newcommand{\listx}{x^{1},x^{2},x^{3}}

\newcommand{\xlist}{x^{1},x^{2},x^{3}}
\newcommand{\nlist}{n^{1},n^{2},n^{3}}
\newcommand{\nlistb}{n^{1},n^{2}}
\newcommand{\listxhat}{\hat{x}^{1},\hat{x}^{2},\hat{x}^{3}}
\newcommand{\xhatlist}{\hat{x}^{1},\hat{x}^{2},\hat{x}^{3}}
\newcommand{\qlist}{q^{1},q^{2},q^{3}}
\newcommand{\plist}{p_{1},p_{2},p_{3}}
\newcommand{\hatnlist}{\hat{n}^{1},\hat{n}^{2},\hat{n}^{3}}
\newcommand{\ksquared}{(k_{1})^{2}+(k_{2})^{2}+(k_{3})^{2}}

\newcommand{\bp}{\bm{P}}
\newcommand{\bq}{\bm{Q}}
\newcommand{\bpsi}{\bm{\psi}}
\newcommand{\romansubs}{\renewcommand{\theequation}{\theparentequation \roman{equation}}}

\begin{document}

\pagestyle{fancy}
\fancyhead{} 
\fancyhead[OR]{\thepage}
\fancyhead[OC]{{\small{DISCRETE PHASE SPACE QUANTUM MECHANICS}}}
\fancyfoot{} 
\renewcommand\headrulewidth{0.5pt}
\addtolength{\headheight}{2pt} 

\title{{\small {\bf{DISCRETE PHASE SPACE: QUANTUM MECHANICS AND NON-SINGULAR POTENTIAL FUNCTIONS}}}}
\author {{\small Anadijiban Das \footnote{das@sfu.ca}} \\
\it{\small Department of Mathematics,  Simon Fraser University} \\
\it{\small Burnaby, British Columbia, V5A 1S6, Canada}
\\[-0.1cm]
\line(1,0){45}\\[0.1cm]
\and
{\small Andrew DeBenedictis \footnote{adebened@sfu.ca}} \\
\it{\small Department of Physics, Simon Fraser University}\\
\it{\small and}\\
\it{\small The Pacific Institute for the Mathematical Sciences,} \\
\it{\small Burnaby, British Columbia, V5A 1S6, Canada }
}
\date{{\small April 7, 2015}}
\maketitle
\begin{abstract}
\noindent The three-dimensional potential equation, motivated by representations of quantum mechanics, is investigated in four different scenarios: (i) In the usual Euclidean space $\mathbb{E}_{3}$ where the potential is singular but invariant under the continuous inhomogeneous orthogonal group $\iothree$. The invariance under the translation subgroup is compared to the corresponding unitary transformation in the Schr\"{o}dinger representation of quantum mechanics. This scenario is well known but serves as a reference point for the other scenarios. (ii) Next, the discrete potential equation as a partial difference equation in a three-dimensional lattice space is studied. In this arena the potential is non-singular but invariance under $\iothree$ is broken. This is the usual picture of lattice theories and numerical approximations. (iii) Next we study the six-dimensional continuous phase space. Here a \ps representation of quantum mechanics is utilized. The resulting potential is singular but possesses invariance under $\iothree$. (iv) Finally, the potential is derived from the discrete \ps representation of quantum mechanics, which is shown to be an \emph{exact} representation of quantum mechanics. The potential function here is both non-singular and possesses invariance under $\iothree$, and this is proved via the unitary transformations of quantum mechanics in this representation.
\end{abstract}
\rule{\linewidth}{0.2mm}
\vspace{-1mm}

\section{{Introduction}}\label{sec:intro}
It is well known that quantum mechanics may be represented in terms of \ps variables \cite{ref:psbook1}. This formulation of quantum mechanics has a history dating back to Wigner \cite{ref:wigner} and Weyl \cite{ref:weyl}. The analog of the ``wave function'' in this formulation is a probability distribution function in \ps known as the Wigner function. It is also known that one may describe quantum mechanics in terms of discrete quantities \cite{ref:matrixmech}, \cite{ref:matrixmech2}. Finally, the discrete \ps representation of quantum mechanics has been established in references \cite{ref:dpqm1} and \cite{ref:dpqm2}. The generalization of this representation to relativistic quantum mechanics and field theory has been presented in \cite{ref:drqm1}, \cite{ref:drqm2}, and \cite{ref:drqm3}. In \cite{ref:drqm3} quantum electrodynamics was formulated in the arena of three-dimensional discrete \ps and continuous time. (Such mixing of discrete and continuous operations is mathematically rigorous and has a long history. For example, see \cite{ref:semidisc1}-\cite{ref:discprop}.\footnote{The particular representation in \cite{ref:drqm1}-\cite{ref:drqm3} also respects Lorentz symmetry, as was shown in \cite{ref:drqm2}. One confusion comes from the fact that it is tempting to think that the spacetime is discrete in these theories, but it is not.})  The great advantage of this representation is that the S-matrix elements turn out to be \emph{divergence free}. The mathematical reason for the elimination of the divergences is the fact that partial difference-differential equations representing free fields have non-singular Green's functions. This discrete \ps representation is relatively new, and has not been studied in any detail in the important physical arenas of non-relativistic quantum mechanics and potential theory, save for specifically the harmonic oscillator \cite{ref:dpqm1}. This is the primary motivation for the current study.

In this paper we concentrate on the partial difference equations representing the potential equation in the three-dimensional discrete \ps (which is a subset of the six-dimensional continuous \ps), and study the resulting non-singular potential, which replaces the usual singular Coulomb potential. Before deriving the result in discrete \ps we discuss briefly in section \ref{sec:potfunE3} the usual elliptic potential equation in the differentiable manifold $\mathbb{E}_{3}$. The Green's functions and the potential function $V_{0}(x^{1},\,x^{2},\,x^{3})$ exhibit well known singularities. Outside the singularity, the potential function is invariant under the continuous group $\mathcal{I}O(3)$. Although this fact is well known, we will utilize a proof of this invariance from the unitary transformation of position and momentum operators, $Q^{j}$ and $P_{k}$, in quantum mechanics according to the Schr\"{o}dinger representation. This is done as a reference for the study of quantum mechanics and potential theory in the other representations in subsequent sections. 

In section \ref{sec:discpot} we touch upon discrete potential theory \cite{ref:duffin} in a three-dimensional lattice space. Physical applications of such a discrete potential occur in random walks on a lattice \cite{ref:discrw} and in finite electrical networks \cite{ref:discelec}. The representation here is also often utilized in numerical approximations. This discrete potential is non-singular. However, it does not admit invariance under the continuous group $\iothree$. Also, this difference Laplacian is \emph{not} capable of providing an exact representation of quantum mechanics.

In section \ref{sec:psqm} we deal with the representation of quantum mechanics in the six-dimensional continuous \ps manifold \cite{ref:psbook1}. (Originally, Weyl and Wigner suggested such representations \cite{ref:weyl} \cite{ref:wigner}.) In this section we also formulate the potential equation in continuous \ps, again by using the corresponding quantum mechanics operators $Q^{i}$ and $P_{k}$. The potential here is singular. However, it does admit the invariance under the group $\iothree$ as a subgroup of canonical transformations.

In section \ref{sec:dpsqm} we investigate the potential theory in the new discrete \ps scenario. Firstly, quantum mechanical operators $Q^{i}$ and $P_{k}$ are formulated in this representation. These are provided by new finite difference operators which we denote as $\delta^{ij}\circled$ and $-i\sharpd_{k}$. These are capable of providing \emph{an exact representation} of quantum mechanics, not just a discrete approximation of the continuum theory. The corresponding Green's functions are non-singular and the discrete \ps potential, $\wn$ is analogous to the potential $\vx$ in $\mathbb{E}_{3}$. The potential equation governing the function $\wn$ is proved to be \emph{invariant under the continuous group $\iothree$}. The proof here is also based on the unitary transformation involving the corresponding quantum mechanical operators $Q^{i}$ and $P_{k}$. We also study the particular solution $\wzn$, analogous to the particular potential $\vzx$ of section \ref{sec:potfunE3} (corresponding to the ``source'' point at the origin). It is shown that $\wzn$ is \emph{non-singular} and, when analysis is restricted to an axis, $W_{(0)}(0,0,2n^{3})$ can be expressed in terms of gamma functions. 

We present the variational derivation of second order partial difference equations and the equations for boundary variations in an appendix for completion \cite{ref:drqm1}.

We choose units such that $\hbar=1$. A new constant, $\ell > 0$, will also appear in subsequent calculations, which is a characteristic length. Except in certain instances where it is required for clarity, we set $\ell=1$.

\section{The potential function in the Euclidean space $\mathbb{E}_{3}$}\label{sec:potfunE3}
In this section we briefly review the usual potential function in $\mathbb{E}_{3}$. Most of the analysis in this section is well known but this section serves to set the stage for subsequent results for comparisons.  Before proceeding we establish the notation as follows. Roman indices take values from $\{1,\,2,\,3\}$ and summation over repeated indices is implied. Hatted quantities refer to transformed versions of the corresponding unhatted quantities. The ordered triple $(x^{1},\,x^{2},\,x^{3}) \in \mathbb{R}^{3}$ denotes a Cartesian coordinate chart for $\mathbb{E}_{3}$.  

The potential function is a solution to the well known equation
\begin{equation}
 \nabla^{2}\vx:=\delta^{jk} \frac{\partial^{2}}{\partial x^{j}\partial x^{k}} \vx \equiv \delta^{jk} \partial_{j}\partial_{k} \vx =0\,. \label{eq:e3poteqn}
\end{equation}
A particular Green's function for the elliptic partial differential equation (\ref{eq:e3poteqn}) is provided by
\begin{align}
 G(x^{1},x^{2},x^{3};\hat{x}^{1},\hat{x}^{2},\hat{x}^{3})=&\frac{1}{(2\pi)^{3}} \int_{-\infty}^{\infty}\int_{-\infty}^{\infty}\int_{-\infty}^{\infty} \frac{\exp\left[ik_{j}\left(x^{j}-\hat{x}^{j}\right)\right]}{\left[(k_{1})^{2} + (k_{2})^{2} + (k_{3})^{2}\right]} dk_{1}dk_{2}dk_{3} \nonumber \\[0.2cm]
 =&\frac{1}{4\pi} \frac{1}{\left[(x^{1}-\hat{x}^{1})^{2}+(x^{2}-\hat{x}^{2})^{2}+(x^{3}-\hat{x}^{3})^{2}\right]^{1/2}} \,. \label{eq:e3gf}
\end{align}
In electrostatic phenomena, the physical interpretation of (\ref{eq:e3gf}) is that it represents the Coulomb potential at field point $(x^{1},x^{2},x^{3})$ due to a positive unit electric charge located at source point $(\hat{x}^{1},\hat{x}^{2},\hat{x}^{3})$ (in Heaviside-Lorentz units).

By taking the ``source point'' at the origin, we can write a particular solution of the potential equation as
\begin{subequations}
\romansubs
 \begin{align}
\vzx:=G(\listx; 0,0,0)=&\frac{1}{4\pi} \frac{1}{\sqrt{(x^{1})^{2}+(x^{2})^{2}+(x^{3})^{2}}} \nonumber \\[0.2cm]
 & \mbox{for } (x^{1})^{2}+(x^{2})^{2}+(x^{3})^{2} > 0\,, \label{eq:e3potzero1} \\[0.2cm]
\lim_{\mbox{\tiny{$|x^{j}|\rightarrow \infty$}}} \left|\vzx\right| =&0\,, \label{eq:e3potzero2} \\[0.2cm]
\lim_{\mbox{\tiny{$(\listx)\rightarrow (0,0,0)$}}} \left|\vzx\right| \rightarrow& \infty\,. \label{eq:e3potzero3}
\end{align}
\end{subequations}
The equation (\ref{eq:e3potzero3}) demonstrates the well known fact that the above potential diverges in the limit $(\xlist)\rightarrow (0,0,0)$.

The potential equation (\ref{eq:e3poteqn}) may be variationally derived from an appropriate action (the Dirichlet integral \cite{ref:candh}, \cite{ref:kellogg}):
\begin{subequations}
 \romansubs
\begin{align}
J[V]:=&\int_{\overline{D}_{3}} dx^{1}dx^{2}dx^{3}\;\;L(y;y_{1},y_{2},y_{3})_{\mbox{\tiny{$|y=V(...), y_{j}=\partial_{j}V(...)$}}} \,, \label{eq:e3action}\\[0.2cm]
L(y;y_{1},y_{2},y_{3})_{...}:=& \frac{1}{2}\left(\delta^{jk}y_{j}y_{k}\right)_{|..} \,. \label{eq:e3lag}
\end{align}
\end{subequations}
Here, $\overline{D}_{3}$ is a closed, simply-connected, proper subset of $\mathbb{E}_{3}$. The varied function is chosen to be 
\begin{equation}
\hat{y}:=\vx +\varepsilon h(\xlist)\,, \label{eq:e3variedfun}
\end{equation}
with $\varepsilon \neq  0$ and $h$ being an arbitrary bounded function of class $C^{2}(\overline{D}_{3}; \mathbb{R})$. The variational principle applied to (\ref{eq:e3action}) using (\ref{eq:e3lag}) implies that
\begin{subequations}
 \romansubs
\begin{align}
 &\delta^{jk}\partial_{j}\partial_{k}\vx  =0\,, \label{eq:e3eleqn} \\[0.1cm]
 \mbox{and} \nonumber \\[0.1cm]
 &\int_{\partial\overline{D}_{3}} \left[  h(\xlist) \delta^{jk} \partial_{k}\vx n_{j}(\xlist)\right]_{|...} d^{2}s =0\,. \label{eq:e3bt}
\end{align}
\end{subequations}
The equation (\ref{eq:e3eleqn}) is the Euler-Lagrange equation in ${D}_{3} \subset \mathbb{R}^{3}$, and the equation (\ref{eq:e3bt}) is the boundary term due to the variation on $\partial \overline{D}_{3}$. In the case of Dirichlet boundary conditions $h(\xlist)_{\partial \overline{D}_{3}}\equiv 0$, and (\ref{eq:e3bt}) is identically satisfied. Moreover, the only prescribed Neumann condition $n^{j}\partial_{j}V(\xlist)_{|\partial\overline{D}_{3}}\equiv 0$ is variationally admissible.

Now we shall briefly deal with the symmetry group admitted by Laplace's equation (\ref{eq:e3poteqn}). It is the six-parameter inhomogeneous orthogonal group $\iothree$. This Lie group is explicitly provided by the transformation
\begin{subequations}
 \romansubs
\begin{align}
\hat{x}^{j}=& c^{j} + r^{j}_{\;k} x^{k}\,, \label{eq:e3transf1}\\[0.2cm]
\delta_{jk}r^{j}_{\;a}r^{k}_{\;b}=&\delta_{ab}\,, \label{eq:e3transf2}\\[0.2cm]
a^{j}_{\;b}r^{b}_{\;k}=& r^{j}_{\;b}a^{b}_{\;k}= \delta^{j}_{\;k}\,, \label{eq:e3transf3}\\[0.2cm]
x^{k}=& a^{k}_{\;j} \left(\hat{x}^{j} - c^{j}\right)\,, \label{eq:e3transf4}
\end{align}
\end{subequations}
with the six independent parameters $c^{j}$ and $r^{j}_{\;k}$ comprising the linear transformation.

The potential function $\vx$, and the potential equation (\ref{eq:e3poteqn}), treated tensorially under (\ref{eq:e3transf1}-iv), transform as
\begin{subequations}
 \romansubs
\begin{align}
 \hat{V}(\listxhat) =& \vx \,,\label{eq:e3pottransf1}\\[0.2cm]
 \hat{\delta}^{ij} \hat{\partial}_{i}\hat{\partial}_{j}\hat{V}(\xhatlist)=& \delta^{ij} a^{b}_{\;i}a^{c}_{\;j} \partial_{b}\partial_{c} \vx = \delta^{bc}\partial_{b}\partial_{c} \vx =0\,. \label{eq:e3pottransf2}
\end{align}
\end{subequations}
The equation (\ref{eq:e3pottransf2}) illustrates the invariance of the potential equation (\ref{eq:e3poteqn}) under the group $\iothree$ expressed in (\ref{eq:e3transf1}-iv). If one sets $a^{j}_{\;k} = \delta^{i}_{\;k}$ and $r^{i}_{\;k}= \delta^{i}_{\;k}$ the transformations comprise the three-parameter Abelian subgroup of translations:
\begin{subequations}
 \romansubs
\begin{align}
 \hat{V}(x^{1}+c^{1},x^{2}+c^{2},x^{3}+c^{3}) =&\, \vx\,,\label{eq:e3pottransl1}\\[0.2cm]
 \mbox{or,}\quad \hat{V}(\xlist) =&\, {V}(x^{1}-c^{1},x^{2}-c^{2},x^{3}-c^{3})\,. \label{eq:e3pottransl2}
\end{align}
\end{subequations}

It is known that the potential function $\vx$ (also called a harmonic function), locally admits the Taylor series expansion in a star-shaped domain of $\mathbb{R}^{3}$ \cite{ref:kellogg}. Therefore, we obtain from (\ref{eq:e3pottransl2}),
\begin{subequations}
 \romansubs
\begin{align}
\hat{V}(\xlist)=&V(\xlist) \nonumber\\[0.1cm]
& +\sum_{j=1}^{\infty} \frac{(-1)^{j}}{j!}  \left[\underset{(i_{1}+...+i_{j}=j)}{\sum_{i_{1}=1}^{3}\cdots\sum_{i_{j}=1}^{3}} \left[c^{i_{1}} \cdots c^{i_{j}}\right] \frac{\partial^{j}}{\partial x^{i_{1}} \cdots \partial x^{i_{j}}} \vx \right]\,, \label{eq:e3taylor1} \\[0.1cm]
 \mbox{or, }\nonumber\\[0.1cm]
 \hat{V}(\xlist)=&\exp\left[-c^{k} \partial_{k}\right] \vx\,, \label{eq:e3taylor2} \\
 \delta^{ab} \partial_{a}\partial_{b} \hat{V}(\xlist) =& \exp\left[-c^{k} \partial_{k}\right] \left[\delta^{ab}\partial_{a}\partial_{b}\vx\right] =0\,. \label{eq:e3taylor3}
\end{align}
\end{subequations}
Thus, we have an alternate proof of the invariance of the potential equation (\ref{eq:e3poteqn}) under the subgroup of translations. This proof has a quantum mechanical aspect to it and will be also useful for subsequent analysis of quantum mechanics in the discrete \ps representation.

In the Schr\"{o}dinger representation of quantum mechanics (also known as wave mechanics) the self-adjoint coordinate and momentum operators acting on Hilbert space vectors are provided by
\begin{subequations}
 \romansubs
\begin{align}
\bm{P}_{k}\vec{\bm{\psi}}:=&-i\partial_{k}\psi(\xlist)\,, \label{eq:schrop}\\[0.2cm]
\bm{Q}^{j}\vec{\bm{\psi}}:=&x^{j}\psi(\xlist)\,, \label{eq:schrox} \\[0.2cm]
\left[\bm{P}_{k}\bm{Q}^{j}-\bm{Q}^{j}\bm{P}_{k}\right]\vec{\bm{\psi}}=&-i\delta^{j}_{\;k}\,\psi(\xlist)\,. \label{eq:schrocomm}
\end{align}
\end{subequations}
Now consider an unusual quantum mechanical equation
\begin{subequations}
 \romansubs
\begin{align}
-\delta^{jk}\bm{P}_{j}\bm{P}_{k}\vec{\bm{\psi}}=&\vec{\bm{0}}\,,\label{eq:psq1}\\[0.1cm]
\mbox{or,}\qquad\qquad \nonumber \\[0.1cm]
\delta^{jk}\partial_{j}\partial_{k}\psi(\xlist)=&0\,.\label{eq:psq2}
\end{align}
\end{subequations}
This corresponds to a free particle with no kinetic or potential energy. The equation (\ref{eq:psq2}) implies that both Re($\psi$) and Im($\psi$) satisfy the potential equation (\ref{eq:e3poteqn}). The above equations can also be understood as the non-relativistic limit of a massless free particle in relativistic quantum mechanics via
\begin{subequations}
 \romansubs
\begin{align}
&\left[\delta^{jk}\bm{P}_{j}\bm{P}_{k}-(\bm{P}_{4})^{2}\right] \vec{\bm{\psi}}=\vec{\bm{0}}\,,\label{eq:relpsq1}\\[0.2cm] 
&\mbox{or }, \nonumber \\[0.1cm]
&-\delta^{jk}\partial_{j}\partial_{k}\psi(\xlist,x^{4}) +\left(\frac{1}{c}\partial_{t}\right)^{2}\psi(\xlist,x^{4})=0\,. \label{eq:relpsq2}
\end{align}
\end{subequations}
Here $c$ is the speed of light and the fourth coordinate, $x^{4}$ is $ct$. In the non-relativistic limit $c \rightarrow \infty$, the above relativistic equation reduces to (\ref{eq:psq2}).

In the Schr\"{o}dinger picture, the transformations of Hilbert space vectors and position and momentum operators, induced by the translation subgroup of $\iothree$ are specified by
\begin{subequations}
 \romansubs
\begin{align}
\hat{\bm{P}}_{j}=&\bm{P}_{j}\,,\,\,\hat{\bm{Q}}^{k}=\bm{Q}^{k}\,, \label{eq:schrotrans1}\\[0.2cm]
\hat{\vec{\bm{\psi}}}=&\exp\left[-ic^{j}\vec{\bm{P}}_{j}\right] \vec{\bm{\psi}}\,.\label{eq:schrotrans2}
\end{align}
\end{subequations}
By equations (\ref{eq:schrop}-iii), the transformations above yield
\begin{equation}
\hat{\psi}(\xlist)=\exp\left[-c^{j}\partial_{j}\right]\psi(\xlist)\,. \label{eq:schrotrans3}
\end{equation}
The equation (\ref{eq:schrotrans3}) above is the exact replica of (\ref{eq:e3taylor2}). Hence the equivalence of such transformations on the potential $\vx$ and on the zero-energy non-relativistic quantum mechanical particle. We repeat here that most of the analysis in this section is well known, but it serves as a reference for the analysis of the potential equation and quantum mechanics in discrete representations and \ps . We will also comment on advantages and disadvantages of these other representations in the appropriate sections.

\section{Discrete space and the potential function}\label{sec:discpot}
There is a characteristic length $\ell>0$ here. (Recall that we use units such that $\ell=1$. This is common in numerical analysis.) There are also three discrete variables $n^{j} \in \left\{...,-1,0,1,2...\right\}=:\mathbb{Z}$. We define various partial difference operators as follows: 

The right partial difference operator is defined as
\begin{equation}
\Delta_{j} f(\nlist):=f(\cdots,n^{j}+1,\cdots)-f(\cdots,n^{j},\cdots)\,. \label{eq:rpdo}
\end{equation}
The left difference operator is defined as
\begin{equation}
\Delta^{\prime}_{j} f(\nlist):=f(\cdots,n^{j},\cdots)-f(\cdots,n^{j}-1,\cdots)\,. \label{eq:lpdo}
\end{equation}
The mean partial difference operator is defined as
\begin{equation}
\overline{\Delta}_{j} f(\nlist):=\frac{1}{2}\left[f(\cdots,n^{j}+1,\cdots)-f(\cdots,n^{j}-1,\cdots)\right]\,. \label{eq:mpdo}
\end{equation}
The discrete potential equation is taken to be \cite{ref:duffin}
\begin{equation}
\delta^{jk}\Delta_{j}\Delta^{\prime}_{k} U(\nlist) = 0\,. \label{eq:discpoteqn}
\end{equation}
It is usually defined this way as this combination of discrete derivatives yield a self-adjoint potential difference equation. As well, it is variationally derivable in this form \cite{ref:das1960} and this is the usual Laplacian of discrete numerical analysis.

The partial difference equation in discrete analysis, (\ref{eq:discpoteqn}), is an analogue of the usual potential equation (\ref{eq:e3poteqn}) in the continuum. The discrete potential equation (\ref{eq:discpoteqn}) implies that
\begin{align}
U(\nlist)=&\frac{1}{6} \Big\{\left[U(n^{1}+1,n^{2},n^{3})
+ U(n^{1}-1,n^{2},n^{3})\right]\nonumber \\ 
 &+ \left[U(n^{1},n^{2}+1,n^{3}) + U(n^{1},n^{2}-1,n^{3})\right]\nonumber \\
 &+ \left[U(n^{1},n^{2},n^{3}+1) + U(n^{1}, n^{2},n^{3}-1)\right]\Big\}\,. \label{eq:discpot}
\end{align}
Thus, the value $U(\nlist)$ of the discrete potential (or discrete harmonic function) is the arithmetic mean of its values at the six neighboring points of $(\nlist)$. This property is analogous to the mean-value theorem for the harmonic function $\vx$ of the continuous variables $(\xlist)$ studied in the preceding section. A Green's function for the discrete potential equation (\ref{eq:discpoteqn}) is furnished by
\begin{subequations}
 \romansubs
\begin{align}
&G(\nlist,\hat{n}^{1},\hat{n}^{2},\hat{n}^{3})&\nonumber\\[0.1cm]
&\qquad:=\frac{1}{4(2\pi)^{3}} \int_{-\pi}^{\pi}\int_{-\pi}^{\pi}\int_{-\pi}^{\pi} \frac{\exp\left[ik_{j}\left(n^{j}-\hat{n}^{j}\right)\right]}{\left[\sin^{2}\left(\frac{k_{1}}{2}\right)+\sin^{2}\left(\frac{k_{2}}{2}\right)+ \sin^{2}\left(\frac{k_{3}}{2}\right)\right]}dk_{1}dk_{2}dk_{3}\,, \label{eq:discgf1}\\[0.2cm]
&\delta^{ij}\Delta_{i}\Delta^{\prime}_{j}G(\cdots)= -\delta_{n^{1}\hat{n}^{1}}\delta_{n^{2}\hat{n}^{2}}\delta_{n^{3}\hat{n}^{3}}\,.\label{eq:discgf2}
\end{align}
\end{subequations}
A discrete potential function $U_{(0)}(\nlist)$, analogous to $\vzx$ of (\ref{eq:e3potzero1}) is provided by
\begin{align}
&U_{(0)}(\nlist):=G(\nlist,0,0,0)&\nonumber\\[0.1cm]
&\qquad=\frac{1}{4(2\pi)^{3}} \int_{-\pi}^{\pi}\int_{-\pi}^{\pi}\int_{-\pi}^{\pi} \frac{\exp\left[ik_{j}n^{j}\right]}{\left[\sin^{2}\left(\frac{k_{1}}{2}\right)+\sin^{2}\left(\frac{k_{2}}{2}\right)+ \sin^{2}\left(\frac{k_{3}}{2}\right)\right]}dk_{1}dk_{2}dk_{3}\,. \label{eq:disczeropot1}
\end{align}

An interesting property of $U_{(0)}(\nlist)$ is
\begin{subequations}
\romansubs
\begin{align}
U_{(0)}(0,0,0)=&\frac{1}{2}\left[18+12\sqrt{2}-10\sqrt{3}-7\sqrt{6}\right] \Big\{\left(\frac{2}{\pi}\right) K\left[(2-\sqrt{3})(\sqrt{3}-\sqrt{2})\right] \Big\}^{2}\,, \label{eq:uzero1}\\[0.2cm]
\left|U_{(0)}(0,0,0)\right| <&\;\infty\,. \label{eq:uzero2}
\end{align}
\end{subequations} 
Here, the function $K[\cdots]$ denotes the complete elliptic integral. Thus, the function $U_{(0)}(\nlist)$ in (\ref{eq:disczeropot1}) is everywhere non-singular. For large values of $n^{j}$ this function possesses the following asymptotic behavior
\begin{subequations}
\romansubs
\begin{align}
U_{(0)}(\nlist)=&\frac{1}{4\pi||\bm{n}||} + \frac{1}{32\pi||\bm{n}||^{3}} \Big\{ \frac{5[({n}^{1})^{4}+(n^{2})^{4}+(n^{3})^{4}]}{||\bm{n}||^{4}} - 3\Big\} +\mathcal{O}\left(\frac{1}{||\bm{n}||^{5}}\right)\,, \label{eq:uzeroinf1}\\[0.2cm]
&\lim_{||\bm{n}||\rightarrow\infty} \left| U_{(0)}(\bm{n}) \right|=0\,, \label{eq:uzeroinf2}
\end{align}
\end{subequations}
with $||\bm{n}||:=\sqrt{(n^{1})^{2}+(n^{2})^{2}+(n^{3})^{2}}$. From the expression in braces in (\ref{eq:uzeroinf1}) it can be inferred that the potential violates the symmetry group $O(3)$. Therefore, although this discrete potential remedies the singularity, it is not invariant under the continuous rotation group. This, of course, is due to the fact that the underlying lattice structure of the theory does not possess rotational invariance. The invariance group of the discrete potential equation (\ref{eq:discpoteqn}) is a crystallographic space group \cite{ref:crystalgroup}.

We finish this section with a brief comment that the difference operators in (\ref{eq:rpdo})-(\ref{eq:mpdo}) are not capable of providing an exact representation of quantum mechanics, although an approximation in the form of the discrete Schr\"{o}dinger equation can be provided \cite{ref:discshro}. The discrete Schr\"{o}dinger equation has been shown to to have applicability in tight-binding models for crystal solids \cite{ref:tightbind} and highly localized phases. The new representation of section \ref{sec:dpsqm} yields non-singular Green's functions and an exact representation of the algebra of quantum mechanics.

\section{Continuous phase space and the potential function}\label{sec:psqm}
This section relies heavily on the representations of quantum mechanics. The usual Schr\"{o}dinger representation was briefly introduced in equations (\ref{eq:schrop})-(\ref{eq:schrocomm}). Although not common, quantum mechanics can also be formulated in \ps \cite{ref:psbook1}. We provide here a continuous \ps representation of quantum mechanics in the following:
\begin{subequations}
\romansubs
\begin{align}
 \bm{P}_{j}\vec{\bm{\psi}} :=& \frac{1}{\sqrt{2}}\left(p_{j}-i\partial_{q^{j}}\right)\psi\left(\qlist;\plist\right)\,, \label{eq:ppsqm} \\[0.2cm]
 \bm{Q}^{k} \vec{\bm{\psi}} :=& \frac{1}{\sqrt{2}}\left(q^{k}+i\partial_{p_{k}}\right)\psi\left(\qlist;\plist\right)\,, \label{eq:qpsqm}\\[0.2cm]
 \left[\bm{P}_{j}\bm{Q}^{k} - \bm{Q}^{k}\bm{P}_{j}\right]\vec{\bm{\psi}}=& -i\delta^{k}_{\; j}\,\psi\left(\qlist;\plist\right)\,. \label{eq:comqsqm}
\end{align}
\end{subequations}
The eigenfunctions of the momentum operator are given by:
\begin{subequations}
\romansubs
\begin{align}
 \psi_{(\bm{k})}(\qlist;\plist)=&A(\plist)\exp\left[i(\sqrt{2}k_{j}-p_{j})q^{j}\right]\,, \label{eq:peigpsqm1}\\[0.1cm]
 \mbox{with }& \nonumber\\[0.1cm]
 \bm{P}_{j}\vec{\bm{\psi}}_{(\bm{k})} =& k_{j} \vec{\bm{\psi}}_{(\bm{k})}\,. \label{eq:peigpsqm2}
\end{align}
\end{subequations}
Here, $A(\plist)\neq 0$ is an arbitrary function. Similarly, the eigenfunctions of the position operator are furnished by:
\begin{equation}
 \psi_{(\bm{x})}(\qlist;\plist)=B(\qlist)\exp\left[-i(\sqrt{2}x^{j}-q^{j})p_{j}\right]\,, \label{eq:qeigpsqm}
\end{equation}
with $B(\qlist)\neq 0$ being another arbitrary function.

The potential equation, according to (\ref{eq:psq1}) and (\ref{eq:ppsqm},ii) is provided by
\begin{subequations}
\romansubs
\begin{align}
 -\delta^{jk}\bp_{j}\bp_{k}\vec{\bpsi} =&\vec{\bm{0}}\,, \label{eq:pspoteqn1}\\
 \mbox{or,}\qquad\qquad\qquad &\nonumber \\
 -\frac{1}{2}\left[\delta^{jk}\left(p_{j}-i\partial_{q^{j}}\right)\left(p_{k}-i\partial_{q^{k}}\right)\right]&\Omega(\qlist;\plist) = 0\,. \label{eq:pspoteqn2}
\end{align}
\end{subequations}
A special solution, which is the analogue in this representation of the potential function in (\ref{eq:e3potzero1}-iii) is given by
\begin{subequations}
\romansubs
\begin{align}
 \Omega_{(0)}(\qlist;\plist) =& \frac{1}{(2\pi)^{3})} \int_{-\infty}^{\infty}\int_{-\infty}^{\infty}\int_{-\infty}^{\infty} \frac{\exp\left[i\left(\sqrt{2}k_{j}-p_{j}\right)q^{j}\right]}{\left[(k_{1})^{2}+(k_{2})^{2}+(k_{3})^{2}\right]} dk_{1}dk_{2}dk_{3}\,, \label{eq:omegazero1}\\[0.1cm]
 \mbox{or,}\qquad\qquad&\nonumber\\[0.1cm]
 \Omega_{(0)}(\qlist;\plist) =&\frac{e^{-ip_{j}q^{j}}}{4\sqrt{2}\pi\sqrt{(q^{1})^{2}+(q^{2})^{2}+(q^{3})^{2}}}\qquad \mbox{for } ||\bm{q}||> 0\,,  \label{eq:omegazero2} \\[0.2cm]
 \lim_{||\bm{q}||\rightarrow\infty} & \left|\Omega_{(0)}(\cdots)\right| = 0\,, \label{eq:omegazero3}\\[0.2cm]
 \lim_{||\bm{q}||\rightarrow 0_{+}}  &\left|\Omega_{(0)}(\cdots)\right| \rightarrow \infty \,. \label{eq:omegazero4}
\end{align}
\end{subequations}
Here, we have a singular potential function (as well as a singular Green's function).

Now we shall touch upon the invariance of the potential equation (\ref{eq:omegazero2}). In \ps the symmetry group is the group of canonical transformations, which preserve the canonical symplectic form. The group $\iothree$ is a subgroup of the canonical transformations. 
The required transformation is explicitly furnished by
\begin{subequations}
\romansubs
\begin{align}
 \hat{q}^{j}\,d\hat{p}_{j} + p_{j} dq^{j} = & dS(\hat{p}, q)\,, \label{eq:pscant1} \\[0.2cm]
 S(\hat{p}, q):=& \left(c^{j} + r^{j}_{\;k} q^{k}\right) \hat{p}_{j}\,, \label{eq:pscant2} \\[0.2cm]
 \hat{q}^{j}= \frac{\partial S(\cdot)}{\partial\hat{p}_{j}} =& c^{j} + r^{j}_{\;k} q^{k}\,, \label{eq:pscant3} \\[0.2cm]
 p_{j}=  \frac{\partial S(\cdot)}{\partial{q}^{j}} = &r^{k}_{\;j} \hat{p}_{k}\,, \label{eq:pscant4}\\[0.2cm]
 \hat{\Omega}\left(\hat{\bm{q}};\hat{\bm{p}}\right)=& \Omega\left(\bm{q};\bm{p}\right)\,, \nonumber \\[0.1cm]
\mbox{or, }&  \label{eq:pscant5} \\[0.1cm]
 \hat{\Omega}\left(\bm{q};\bm{p}\right) =& \Omega\left[a^{k}_{\;j}\left(q^{j}-c^{j}\right) ; r^{k}_{\;j} p_{k}\right]\,, \nonumber
\end{align}
\end{subequations}
($a^{k}_{\;j}$ being the inverse transformation of $r^{k}_{\;j}$ and the parameters are in accordance with equations (\ref{eq:e3transf1}-iv)). Note that the particular transformation in (\ref{eq:pscant3}) exactly corresponds to the group $\iothree$ of equation (\ref{eq:e3transf1}). Therefore, the group $\iothree$ is also a symmetry group of the \ps potential $\Omega$.

\section{Discrete phase space: quantum mechanics and the potential function}\label{sec:dpspf}\label{sec:dpsqm}
 It is well known that in the one-dimensional idealized oscillator (reinstating the constants $\hbar$ and $\ell > 0$), the energy eigenvalues are provided by
\begin{align}
 &\frac{1}{2}\left[\frac{\ell^{2}}{\hbar^{2}}\bm{P}^{2}+\frac{1}{\ell^2}\bm{Q}^{2}\right]\vec{\bm{\psi}}_{(N)}=\left(N+\frac{1}{2}\right)\vec{\bm{\psi}}_{(N)}\,, \label{eq:hoeval}\\[0.2cm]
 &N  \in  \left\{0,1,2,\cdots\right\}\,. \nonumber 
\end{align}
Note that the operator $\frac{\ell^{2}}{\hbar^{2}}\bm{P}^{2}+\frac{1}{\ell^2}\bm{Q}^{2}$ has discrete eigenvalues in spite of $\bm{P}$ and $\bm{Q}$ both possessing continuous spectra. The equation (\ref{eq:hoeval}), interpreted phenomenologically, yields the discretization of the phase plane as  
\begin{align}
&\frac{1}{2}\left[\frac{\ell^{2}}{\hbar^{2}} \left(p\right)^{2} + \frac{\left(x\right)^{2}}{\ell^{2}}\right]=n+\frac{1}{2} \,. \label{eq:nplane} \\[0.2cm]
 &n  \in  \left\{0,1,2,\cdots\right\}\,. \nonumber
\end{align} 
The equation (\ref{eq:nplane}) represents denumerably infinite numbers of confocal ellipses in the background of the continuous $p-x$ phase plane. Discrete \ps in this section comprises of three such equations
\begin{align}
 &\frac{1}{2}\left[\frac{\ell^{2}}{\hbar^{2}} \left(p_{j}\right)^{2} + \frac{\left(x^{j}\right)^{2}}{\ell^{2}}\right]-\frac{1}{2}=n^{j}\,. \label{eq:ndef} \\[0.2cm]
 &n^{j}  \in  \left\{0,1,2,\cdots\right\}\,. \nonumber
\end{align} 
In other words, the particle's accessible \ps is discrete in this case in such a way that the above combination of $x^{j}$ and $p_{j}$ is integer valued \cite{ref:drqm1}. In much of the remaining section we again choose units so that the characteristic length $\ell=1$ and $\hbar=1$.

Previously we had defined three partial difference operators in (\ref{eq:rpdo}), (\ref{eq:lpdo}), and (\ref{eq:mpdo}). Here we need to introduce two more partial difference operators in the following. The weighted mean difference operator is
\begin{equation}
 \Delta^{\#}_{j} f(\nlist):=\frac{1}{\sqrt{2}} \left[ \sqrt{n^{j}+1} f(..,n^{j}+1,..) -\sqrt{n^{j}} f(..,n^{j}-1,..)\right]\,, \label{eq:mdo}
\end{equation}
and the weighted average difference operator is
\begin{equation}
 \circled_{j} f(\nlist):=\frac{1}{\sqrt{2}} \left[ \sqrt{n^{j}+1} f(..,n^{j}+1,..) +\sqrt{n^{j}} f(..,n^{j}-1,..)\right]\,, \label{eq:wado}
\end{equation}
Moreover $f(\nlist):=0$ for any $n^{j}<0$. The physical significance of the operators (\ref{eq:mdo}) and (\ref{eq:wado}) is that they provide an \emph{exact} discrete representation of quantum mechanics via
\begin{subequations}
\romansubs
\begin{align}
 \bp_{j}\vec{\bpsi}:=& -i \Delta^{\#}_{j}\, \psi(\nlist)\,, \label{eq:dpsmom} \\[0.2cm]
 \bq^{k}\vec{\bpsi}:=& \delta^{kl}\circled_{l}\, \psi(\nlist) \,, \label{eq:dpspos} \\[0.2cm]
 \left[\bp_{j}\bq^{k} - \bq^{k}\bp_{j}\right] \vec{\bpsi} =& -i \delta^{k}_{\; j}\, \psi(\nlist) \,. \label{eq:dpscom}
\end{align}
\end{subequations}

Now, we denote a Hermite polynomial, $H_{n}(k)$ and define a related polynomial as follows:
\begin{subequations}
\romansubs
\begin{align}
&\xi_{n^{j}}(k_{j}):= \frac{(i)^{n^{j}} e^{-(k_{j})^{2}} H_{n^{j}}(k_{j})}{\pi^{1/4} 2^{n^{j}/2} \sqrt{n_{j}!}}\,; \label{eq:aoherm1} \\[0.1cm]
&\mbox{for } n^{j} \in \left\{0,1,2,...\right\}\,,\nonumber \\[0.1cm]
&\int_{-\infty}^{\infty} \overline{\xi}_{n^{j}} {\xi}_{\hat{n}^{j}}\,dk_{j} =  \delta_{n^{j}\,\hat{n}^{j}}\,,  \label{eq:aoherm2} \\[0.1cm]
&-i \sharpd_{j} \xi_{n^{j}} (k_{j}) = k_{j} \xi_{n^{j}} (k_{j})\,.  \label{eq:aoherm3}
\end{align}
\end{subequations}
(Here the index $j$ is \emph{not} summed.) We can infer from (\ref{eq:aoherm2}, iii) that $\xi_{n^{j}}(k_{j})$ is the normalized eigenfunction of the momentum operator $\bp_{j}\equiv -i\sharpd_{j}$.

\begin{figure}[ht]
\begin{center}
\includegraphics[bb=0 0 555 300,clip, scale=0.7, keepaspectratio=true]{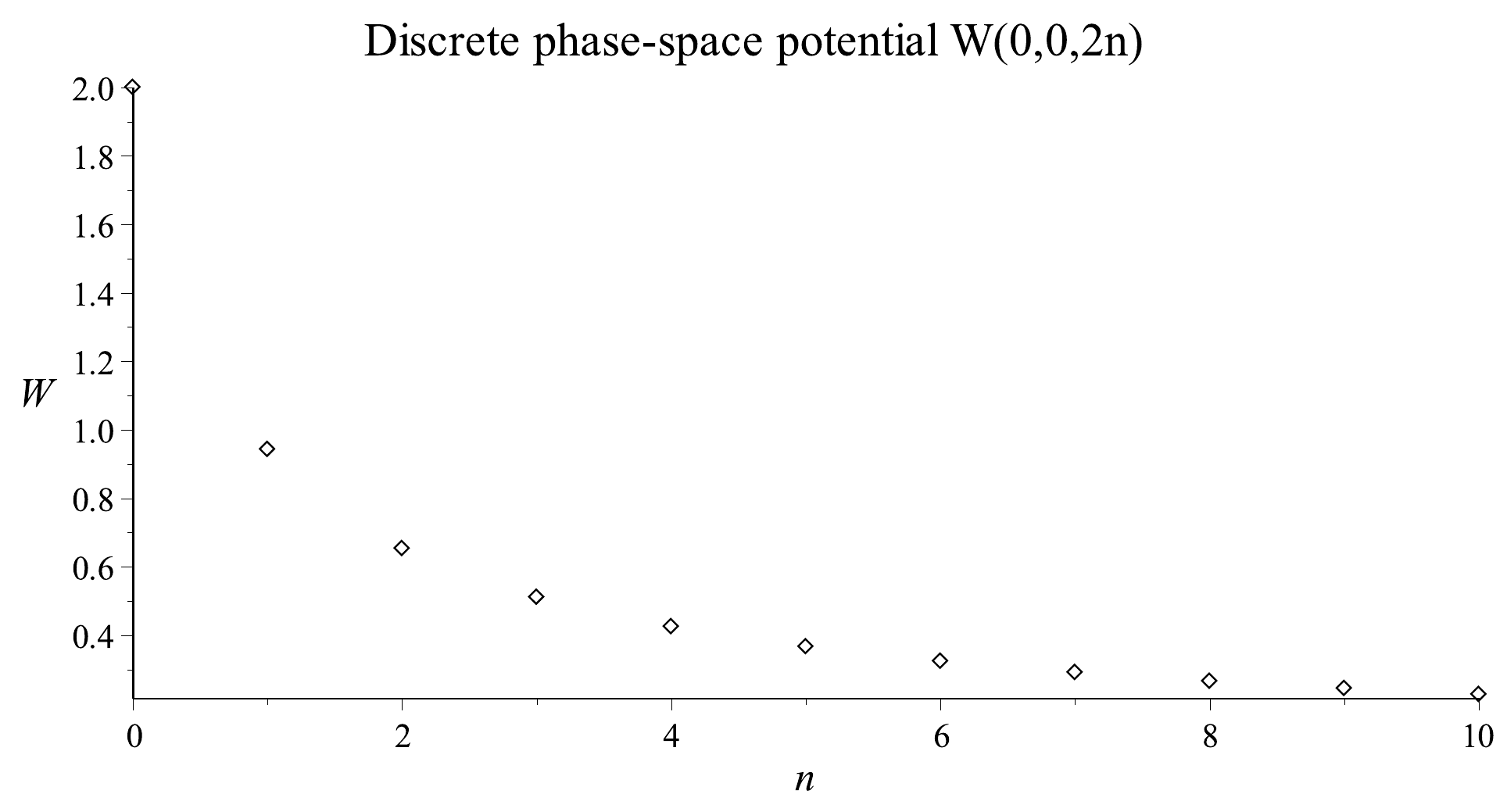}
\caption{{\small The discrete \ps potential $W(0,0,2n)$ for $\ell=0.1$. Note that this potential is finite everywhere and has a value of $2$ at the origin.}}
\label{fig:w}
\end{center}
\end{figure}

The discrete \ps potential equation, analogous to the equations (\ref{eq:e3poteqn}), is chosen to be

\begin{equation}
 -\delta^{jk}\bp_{j}\bp_{k} \vec{\bm{W}} =\delta^{jk} \sharpd_{j}\sharpd_{k} W(\nlist) =0\,. \label{eq:dpspoteqn}
\end{equation}
A Green's function for the partial difference equation (\ref{eq:dpspoteqn}), which is analogous to that in (\ref{eq:e3gf}) is provided by:
\begin{subequations}
\romansubs
\begin{align}
 G(\nlist;\hatnlist):=&\iiint\limits_{-\infty}^{\;\;\;\infty} \frac{\left[\overset{3}{\underset{j=1}{\prod}}\overline{\xi}_{n^{j}}(k_{j}) \xi_{\hat{n}^{j}}(k_{j})\right]}{\left[(k_{1})^{2}+(k_{2})^{2}+(k_{3})^{2}\right]} dk_{1}dk_{2}dk_{3} \,, \label{eq:dspgf1} \\[0.2cm]
 \delta^{jk}\sharpd_{j}\sharpd_{k} G(\nlist;\hatnlist) =& -\delta_{n^{1}\hat{n}^{1}}\delta_{n^{2}\hat{n}^{2}}\delta_{n^{3}\hat{n}^{3}}\,, \label{eq:dspgf2} \\[0.2cm]
 G(0,0,0;0,0,0) = & 2\,. \label{eq:dspgf3}
\end{align}
\end{subequations}
In analogy with (\ref{eq:e3potzero1}), we present the following potential from (\ref{eq:dspgf1}) in this representation:
\begin{align}
 &W_{(0)}(\nlist) = G(\nlist;0,0,0) \nonumber \\[0.2cm]
 &= \Scale[0.99]{\frac{(i)^{n^{1}+n^{2}+n^{3}}}{\pi^{3/2}\,2^{(n^1+n^2+n^3)/2} \left(n^{1}!\,n^{2}!\,n^{3}!\right)^{1/2}} \iiint\limits_{-\infty}^{\;\;\;\infty} \frac{e^{-(k_{1}^{2}+k_{2}^{2}+k_{3}^{2})}\left[\overset{3}{\underset{j=1}{\prod}} H_{n^{j}}(k_{j})\right]}{\left[(k_{1})^{2}+(k_{2})^{2}+(k_{3})^{2}\right]} dk_{1}dk_{2}dk_{3}}\,. \label{eq:dspgzero}
\end{align}
Restricting our analysis along the third axis for the moment\footnote{We will show later that the potential is rotationally invariant and so there is little loss of generality with this restriction and is done for mathematical convenience.} $(0,0,2n)$ admits the closed form expression:
\begin{subequations}
\romansubs
\begin{align}
W_{(0)}(0,0,2n)=& \frac{(-1)^{n}}{\pi^{3/2} 2^{n} \sqrt{2n!}} \iiint\limits_{-\infty}^{\;\;\;\infty} \frac{e^{-\left(\ksquared\right)} H_{2n}(k_{3})}{\left[\ksquared\right]} dk_{1}dk_{2}dk_{3}\,, \label{eq:wzeroz1}\\[0.2cm]
=& \frac{2^{n+1} n!}{(2n+1)\sqrt{(2n)!}} > 0\,, \label{eq:wzeroz2}\\[0.2cm]
&\left[\frac{W_{(0)}(0,0,2n+2)}{W_{(0)}(0,0,2n)}\right] = \left(\frac{2n+2}{2n+3}\right)\sqrt{\frac{2n+1}{2n+2}} < 1\,, \label{eq:wzeroz3}\\[0.2cm]
&\qquad W_{(0)}(0,0,0) =  2\,.\label{eq:wzeroz4}
\end{align}
\end{subequations}
Thus, $\left\{W_{(0)}(0,0,2n)\right\}_{0}^{\infty}$ is a positive-valued, monotone decreasing sequence. Now we shall explore the asymptotic behavior in the limit ${n\rightarrow \infty}\;\,W_{(0)}(0,0,2n)$. For that purpose, we express (\ref{eq:wzeroz2}) as 
\begin{equation}
W_{(0)}(0,0,2n)=\frac{2^{n+1}\,\Gamma(n+1)}{(2n+1)\sqrt{\Gamma(2n+1)}}\,. \label{eq:gammaw}
\end{equation}
The asymptotic expression for the gamma function is provided by
\begin{equation}
\Scale[0.97]{\Gamma(n)=\sqrt{2\pi} (n)^{n-\frac{1}{2}} e^{-n} \left[1 + \frac{1}{12n}+\frac{1}{288n^{2}} -\frac{139}{5184n^{3}} -\frac{571}{2488320n^{4}} +\mathcal{O}\left(\frac{1}{n^{5}}\right)\right]}\,. \label{eq:asympgam}
\end{equation}
Therefore, equations (\ref{eq:wzeroz2}), (\ref{eq:gammaw}), and (\ref{eq:asympgam}) imply that
\begin{subequations}
\romansubs
\begin{align}
W_{(0)}(0,0,2n)=&\frac{\pi^{1/4}}{\sqrt{e}}\Bigg\{ \frac{1}{n^{3/4}} \left[\frac{1+\frac{1}{n}}{1+\frac{1}{2n}}\right]^{n} \left[\frac{(1+\frac{1}{n})^{1/2}}{(1+\frac{1}{2n})^{5/4}}\right]\nonumber\\[0.2cm]
&\times\left[\frac{1+\frac{1}{12(n+1)} +\mathcal{O}\left(\frac{1}{(n+1)^{2}}\right)}{1+\frac{1}{24(2n+1)} +\mathcal{O}\left(\frac{1}{(2n+1)^{2}}\right)}\right]\Bigg\}\,,\label{eq:dpswasymp1} \\[0.2cm] 
\lim_{n\rightarrow\infty}W_{(0)}(0,0,2n)=&\frac{\pi^{1/4}}{\sqrt{e}} \bigg\{ 0\cdot\sqrt{e}\cdot 1 \cdot 1 \bigg\}=0\,. \label{eq:dpswasymp2}
\end{align}
\end{subequations}
We plot $W(0,0,2n)$ in figure \ref{fig:w} for values of $n$ near the origin.

Now we want to compare the continuous \ps potential $\Omega_{0}(\qlist;\plist)$ to the discrete \ps potential $W_{(0)}(0,0,2n)$. Note that $\Omega_{0}(\qlist;\plist)$ is a complex potential, and hence we really wish to compare the real, monotone $W_{(0)}(0,0,2n)$ with the quantity $\sqrt{\left|\Omega_{0}(0,0,q^{3};\plist)\right|^{2}}$. This comparison is also equivalent to comparing $W_{(0)}(0,0,2n)$ to the potential $V_{(0)}(0,0,x^{3})=(1/4\pi)1/|x^{3}|$ noting that the $p_{j}$ dependence drops out in the magnitude $|\Omega_{0}(0,0,q^{3};\plist)|^{2}$. Hence, $\sqrt{\left|\Omega_{0}(0,0,q^{3};\plist)\right|^{2}}$ and $V_{(0)}(0,0,x^{3})$ differ only by a numerical factor of $1/\sqrt{2}$. 

\begin{figure}[ht]
\begin{center}
\includegraphics[bb=0 0 555 320,clip, scale=0.7, keepaspectratio=true]{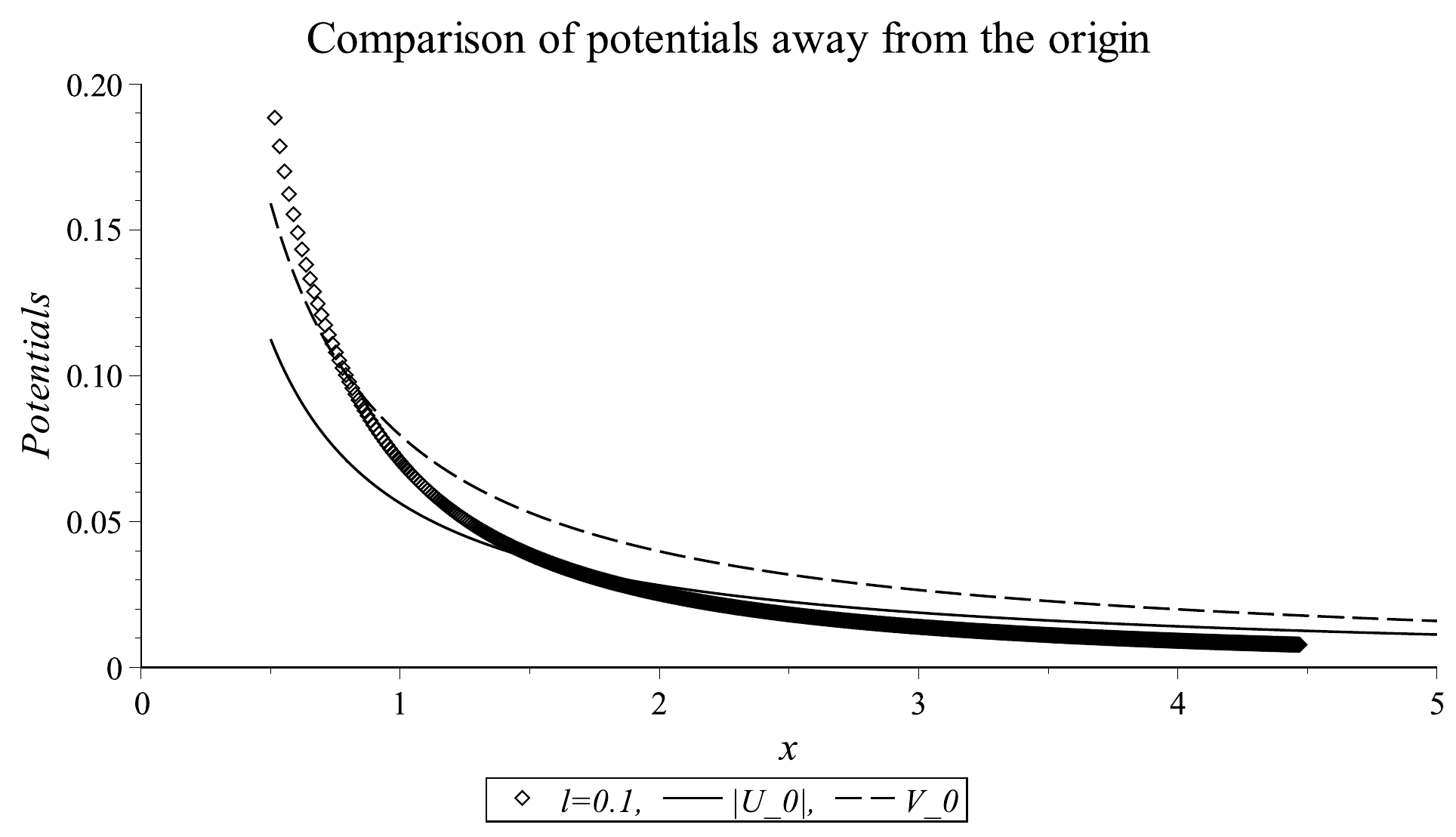}
\caption{{\small A comparison of the potentials $V_{0}$ (dashed), $|U_{(0)}|$ (black) and $W_{(0)}$ (diamonds, for $\ell=0.1$) away from the origin. The discrete \ps potential is actually finite at the origin (see main text and previous plot) whereas the other two potentials diverge at the origin.}}
\label{fig:pot_comp}
\end{center}
\end{figure}

From (\ref{eq:ndef}) we can use the approximation
\begin{equation}
n^{3} = (1/2\ell^{2})(x^{3})^{2}-1/2 +\mathcal{O}(\ell^{2}) \label{eq:napprox}
\end{equation}
for extremely small values of $\ell$. Specifically $\ell$ is assumed small in comparison to $\hbar$, perhaps even as small as the Planck length. Some caution must be applied in the comparison as the $\ell$ and $\hbar$ are not of the same dimension and hence the approximation (\ref{eq:napprox}) only really becomes valid for large values of $x$ and provided the momentum is not large. In this limit the discrete \ps potential function essentially goes over to a configuration space potential function, making the comparison meaningful. Hence we limit our analysis away from the origin. (We have already shown in (\ref{eq:wzeroz4}) and figure \ref{fig:w} that the potential $W_{(0)}(\nlist)$, without the approximations, is finite at the origin.) This comparison is done in figure \ref{fig:pot_comp} for the various potentials. 

The most general transformation of the scalar field $W(\nlist)$ under the six-parameter (proper) continuous group $\iothreeplus$ is given by\footnote{Here, $O^{+}(3)$ denotes the proper subgroup (rotation group) of the orthogonal group $O(3)$.}
\begin{subequations}
\romansubs
\begin{align}
\hat{W}(\nlist)=&\exp\left\{-c^{j}\sharpd_{j} - \frac{\omega^{jk}}{4} \left[\circled_{j}\sharpd_{k} -\circled_{k}\sharpd_{j} + \sharpd_{k}\circled_{j} -\sharpd_{j}\circled_{k}\right]\right\}W(\nlist)\,, \label{eq:invdsw1}\\[0.2cm]
\omega^{kj}\equiv& -\omega^{jk}\,.\label{eq:invdsw2}
\end{align}
\end{subequations}
The translation subgroup is parameterized by the continuous parameters $c^{j}$ and rotations are encoded in the continuous parameters $\omega^{jk}$. The discrete \ps potential equation remains \emph{invariant} since the generators
$\sharpd_{j}$and\allowbreak \\$\left[\circled_{j}\sharpd_{k} -\circled_{k}\sharpd_{j} + \sharpd_{k}\circled_{j} -\sharpd_{j}\circled_{k}\right]$ of the group $\iothreeplus$ commute with the Casimir invariant operator $\delta^{jk}\sharpd_{j}\sharpd_{k}$. Therefore the potential is also invariant under the continuous group $\iothree$. 

Finally, we shall investigate the variational derivation of the discrete \ps potential (\ref{eq:dpspoteqn}). Consider a class of action sum
\begin{equation}
 J[f]:=\sum_{n^{1}=N^{1}_{1}}^{N^{1}_{2}}\sum_{n^{2}=N^{2}_{1}}^{N^{2}_{2}}\sum_{n^{3}=N^{3}_{1}}^{N^{3}_{2}}L(y;y_{1},y_{2},y_{3})_{y=f(\bm{n}),y_{j}=\sharpd_{j}f(\bm{n})}\,. \label{eq:discact}
\end{equation}

The variational principle states:
\begin{equation}
 \lim_{\varepsilon \rightarrow 0} \left\{ \frac{J[f+\varepsilon h] - J[f]}{\varepsilon} \right\} =0\,. \label{eq:actextrem}
\end{equation}
The equation (\ref{eq:actextrem}) implies the following Euler-Lagrange equations:
\begin{equation}
 \frac{\partial L(\cdots)}{\partial y}_{|\cdots} -\sharpd_{j}\left[\frac{\partial L(\cdots)}{\partial y_{j}}\right]_{\cdots} =0 \label{eq:discELeqn}
\end{equation}
and boundary terms. The derivation of (\ref{eq:actextrem}) and the explicit boundary terms will be shown in the Appendix. In the case of the discrete \ps potential equation, the appropriate Lagrangian is furnished by
\begin{equation}
 L(y;y_{1},y_{2},y_{3})_{|y=W(\bm{n}), y_{j}=\sharpd_{j}W(\bm{n})}:=\frac{1}{2}\delta^{jk} \sharpd_{j}W(\bm{n})\sharpd_{k}W(\bm{n})\,. \label{eq:dpslag}
\end{equation}
(Recall that it is the momentum operator which yields the potential equation Laplacian and hence $\sharpd_{j}$ (but not $\circled_{j}$) appears here.)

\section{Concluding remarks}
In this paper we have studied potential theory in discrete \ps. The Green's functions are non-singular as are their associated potentials. Unlike the usual discrete theory which arises as an approximation of continuous systems, these Green's functions have been shown here to be rotationally covariant. As well, the operators giving rise to these Green's functions provide an \emph{exact} representation of quantum mechanics. This provides a novel representation for quantum mechanics. It is known that singular Green's functions, for example, lead to divergences in $S$-matrix elements in relativistic quantum field theory. With the relativistic version of the representation presented here, the corresponding $S$-matrix elements have been previously shown to be finite \cite{ref:drqm3}.


\appendix
\section{APPENDIX - Variational principle in discrete \ps}
We will consider the two-dimensional lattice function $f(\nlistb)$ for the sake of simplicity as generalizations to higher dimensions is straight forward. We use capital Roman indices to take values from $\{1,2\}$. The action functional is defined by
\begin{equation}
 J[f]:=\sum_{n^{1}=N^{1}_{1}}^{N^{1}_{2}}\sum_{n^{2}=N^{2}_{1}}^{N^{2}_{2}}L(y;y_{1},y_{2})_{y=f(\bm{n}),y_{A}=\sharpd_{A}f(\bm{n})}\,. \label{eq:Aact} 
\end{equation}
Subsequently, we denote\;\; $|_{(\nlistb)} \equiv |_{y=f(\nlistb),\sharpd_{A}f(\nlistb)}$\; for brevity.

The variational principle states that
\begin{align}
 0=&\lim_{\varepsilon \rightarrow 0} \left\{ \frac{J[f+\varepsilon h] - J[f]}{\varepsilon}\right\} \nonumber \\[0.1cm]
 =&\sum_{n^{1}=N^{1}_{1}}^{N^{1}_{2}}\sum_{n^{2}=N^{2}_{1}}^{N^{2}_{2}}\left\{\frac{\partial L(\cdots)}{\partial {y}}_{|(\nlistb)} h(\nlistb) +\frac{\partial L(\cdots)}{\partial y_{A}}_{|(\nlistb)} \sharpd_{A}h(\nlistb)\right\}\,. \label{eq:Avarprinc}
\end{align}
We now use the following relation \cite{ref:drqm1} 
\begin{align}
&\phi(\nlistb) \sharpd_{A}h(\nlistb) + h(\nlistb)\sharpd_{A}\phi(\nlistb) \nonumber \\[0.1cm]
&=\Delta_{A}\left\{\frac{\sqrt{n_{A}}}{2} \left[\phi(\nlistb) h(.,n^{A}-1,.) + h(\nlistb) \phi(.,n^{A}-1,.)\right]\right\}\,, \label{eq:Areln}
\end{align}
where $\Delta_{A}$ has been defined in (\ref{eq:rpdo}).
(In the above equation the summation convention has been suspended.)

\begin{figure}[ht]
\begin{center}
\includegraphics[bb=0 0 925 605,clip, scale=0.4, keepaspectratio=true]{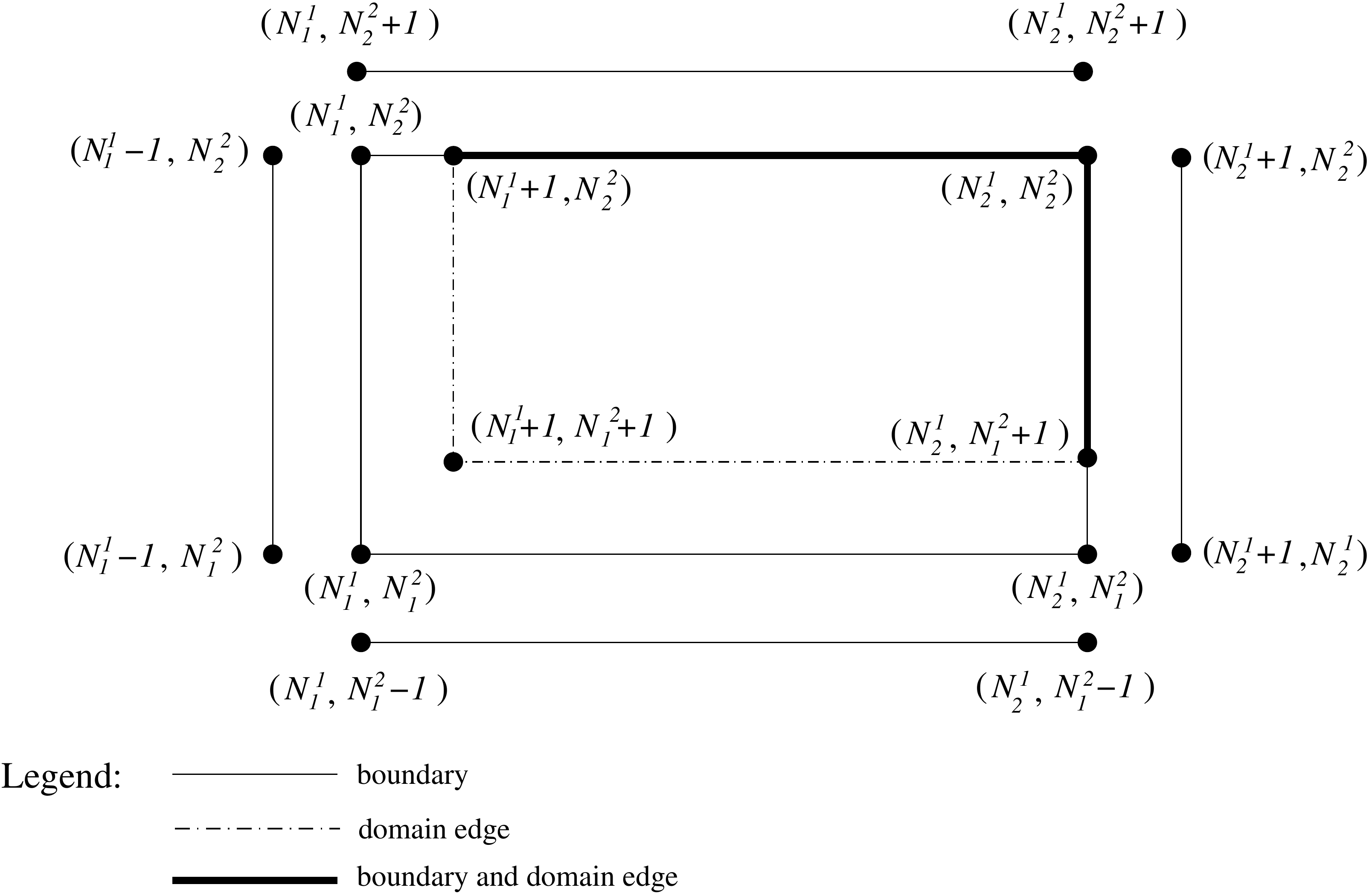}
\caption{{\small The domain and boundary for the discrete variational principle in a two dimensional lattice space.}}
\label{fig:boundaries}
\end{center}
\end{figure}

Next we utilize the discrete Gauss' theorem \cite{ref:duffin} for a discrete vector field $j^{A}(n^{1},n^{2})$:
\begin{align}
 &\sum_{n^{1}=N^{1}_{1}}^{N^{1}_{2}}\sum_{n^{2}=N^{2}_{1}}^{N^{2}_{2}} \Delta_{A}\, j^{A}(\nlistb)  \label{eq:Agt} \\[0.1cm]
 &\quad = \sum_{n^{2}=N^{2}_{1}}^{N^{2}_{2}}\left[j^{1}(N^{1}_{2}+1,n^{2})-j^{1}(N^{1}_{1},n^{2})\right] + \sum_{n^{1}=N^{1}_{1}}^{N^{1}_{2}}\left[j^{2}(n^{1},N^{2}_{2}+1)-j^{1}(n^{1},N^{2}_{1})\right]\,. \nonumber
\end{align}
Using (\ref{eq:Areln}) and (\ref{eq:Agt}) in (\ref{eq:Aact}), and utilizing the discrete Dubois-Reymond lemma \cite{ref:candh}, \cite{ref:drqm1} we derive the following:
\begin{subequations}
\romansubs
{\allowdisplaybreaks\begin{align}
&\mbox{The Euler-Lagrange equations:} \nonumber \\[0.2cm]
&\qquad \frac{\partial L(\cdots)}{\partial y}_{|(\nlistb)} -\sharpd_{j}\left[\frac{\partial L(\cdots)}{\partial y_{j}}\right]_{(\nlistb)} =0 \label{eq:AELeqn} \\[0.1cm]
&\mbox{for }\;\; N^{A}_{1} < n^{A} < N^{A}_{2}+1\; ; \nonumber \\[0.2cm]
&\mbox{The boundary terms:} \nonumber \\[0.2cm]
&\sum_{n^{2}=N^{2}_{1}}^{N^{2}_{2}} \left\{ \sqrt{\frac{N^{1}_{2}+1}{2}} \left[\frac{\partial L(\cdots)}{\partial y_{1}}_{|(N^{1}_{2},n^{2})} h(N^{1}_{2}+1,n^{2}) +\frac{\partial L(\cdots)}{\partial y_{1}}_{|(N^{1}_{2}+1,n^{2})} h(N^{1}_{2},n^{2})\right] \right. \nonumber \\[0.1cm]
&-\sqrt{\frac{N^{1}_{1}}{2}} \left[\frac{\partial L(\cdots)}{\partial y_{1}}_{|(N^{1}_{1},n^{2})} h(N^{1}_{1}-1,n^{2}) \right] \nonumber \\[0.1cm]
&+\left.\left[\frac{\partial L(\cdots)}{\partial y}_{|(N^{1}_{1},n^{2})}-\sharpd_{A}  \frac{\partial L(\cdots)}{\partial y_{A}}_{|(N^{1}_{1},n^{2})} -\sqrt{\frac{N^{1}_{1}}{2}}  \frac{\partial L(\cdots)}{\partial y_{1}}_{|(N^{1}_{1}-1,n^{2})}\right] 
h(N^{1}_{1},n^{2})\right\} \nonumber \\[0.1cm]
&\sum_{n^{1}=N^{1}_{1}}^{N^{1}_{2}} \left\{ \sqrt{\frac{N^{2}_{2}+1}{2}} \left[\frac{\partial L(\cdots)}{\partial y_{2}}_{|(n^{1},N^{2}_{2})} h(n^{1},N^{2}_{2}+1) +\frac{\partial L(\cdots)}{\partial y_{2}}_{|(n^{1},N^{2}_{2}+1)} h(n^{1},N^{2}_{2})\right] \right. \nonumber \\[0.1cm]
&-\sqrt{\frac{N^{2}_{1}}{2}} \left[\frac{\partial L(\cdots)}{\partial y_{2}}_{|(n^{1},N^{2}_{1})} h(n^{1},N^{2}_{1}-1) \right] \nonumber \\[0.1cm]
&+\left.\left[\frac{\partial L(\cdots)}{\partial y}_{|(n^{1},N^{2}_{1})}-\sharpd_{A}  \frac{\partial L(\cdots)}{\partial y_{A}}_{|(n^{1},N^{2}_{1})} -\sqrt{\frac{N^{2}_{1}}{2}}  \frac{\partial L(\cdots)}{\partial y_{2}}_{|(n^{1},N^{2}_{1}-1)}\right]
h(n^{1},N^{2}_{1})\right\} = 0\,. \label{eq:Abt}
\end{align}}
\end{subequations}
Note that the discrete domain $D_{2}$ for the validity of the Euler-Lagrange equations is given by
\begin{equation}
D_{2}:=\left\{ (n^{1},n^{2}): N_{1}^{\;1} \leq n^{1} < N_{2}^{\;1}+1, \, N_{1}^{\;2} < n^{2} < N_{2}^{\;2}+1\right\} \,. \label{eq:Adomain}
\end{equation}
However, the variationally admissible boundary of the discrete domain is provided by
\begin{align}
\partial^{\#}D_{2} =&\left\{(n^{1},n^{2}): n^{1}=N_{2}^{\;1}+1, N_{2}^{\;1}, N_{1}^{\;1}-1, N_{1}^{\;1}; N_{1}^{\;2} \leq n^{2} \leq N_{2}^{\;2}\right\} \nonumber \\[0.1cm]
&\cup \left\{(n^{1},n^{2}): N_{1}^{\;1} \leq n^{1} \leq N_{2}^{\;1}; n^{2}=N_{2}^{\;2}+1, N_{2}^{\;2}, N_{1}^{\;2}-1, N_{1}^{\;2}\right\}\, .\label{eq:Avarbound}
\end{align}
The domain $D_{2}$ and the boundary $\partial^{\#}D_{2}$ are depicted in figure \ref{fig:boundaries}.


\linespread{0.6}
\bibliographystyle{unsrt}

\end{document}